\documentclass[a4paper,fleqn]{cas-dc}

\usepackage[square,numbers]{natbib}

\def\tsc#1{\csdef{#1}{\textsc{\lowercase{#1}}\xspace}}
\tsc{WGM}
\tsc{QE}
\tsc{EP}
\tsc{PMS}
\tsc{BEC}
\tsc{DE}

\begin{document}
\let\WriteBookmarks\relax
\def\floatpagepagefraction{1}
\def\textpagefraction{.001}

\shorttitle{Ultrafast nano generation of acoustic waves in water via a single carbon nanotube}

\shortauthors{M. Diego et~al.}

\title [mode = title]{Ultrafast nano generation of acoustic waves in water via a single carbon nanotube\\
\begin{normalsize}
\vspace{0.25cm}
This article was published in \textit{Photoacoustics} 28, 100407 (2022) and may be found at https://doi.org/10.1016/j.pacs.2022.100407
\end{normalsize}
}

%
\author[1]{\color{black}{Michele Diego}}[type=editor,
                        auid=000,bioid=1,
                        prefix=,
                        role=,    orcid=0000-0003-0807-9414]

\cormark[1]
\ead{michele.diego@univ-lyon1.fr}
\author[2,3,4]{\color{black}{Marco Gandolfi}}[type=editor,
                        auid=000,bioid=1,
                        prefix=,
                        role=,
                        orcid=0000-0001-7700-9255]
\cormark[2]
\ead{marco.gandolfi1@unibs.it}
\author[1,5]{\color{black}{Alessandro Casto}}
\author[5]{\color{black}{Francesco Maria Bellussi}}
\author[1]{\color{black}{Fabien Vialla}}
\author[1]{\color{black}{Aur\'elien Crut}}
\author[6,7]{\color{black}{Stefano Roddaro}}
\author[5]{\color{black}{Matteo Fasano}}
\author[1]{\color{black}{Fabrice Vall\'ee}}
\author[1,8]{\color{black}{Natalia Del Fatti}}
\author[1]{\color{black}{Paolo Maioli}}
\author[1]{\color{black}{Francesco Banfi}}
\affiliation[1]{organization={FemtoNanoOptics group, Universit\'e de Lyon, CNRS, Universit\'e Claude Bernard Lyon 1, Institut Lumi\`ere Mati\`ere},
    addressline={10 Rue Ada Byron}, 
    city={Villeurbanne},
    postcode={F-69622}, 
    country={France}}
    
\affiliation[2]{organization={CNR-INO},
    addressline={via Branze 45}, 
    city={Brescia},
    postcode={25123}, 
    country={Italy}}
    
\affiliation[3]{organization={Department of Information Engineering, Universit\`a di Brescia},
    addressline={via Branze 45}, 
    city={Brescia},
    postcode={25123}, 
    country={Italy}}
    
\affiliation[4]{organization={Interdisciplinary Laboratories for Advanced Materials Physics (I-LAMP) and Dipartimento di Matematica e Fisica, Universit\`a Cattolica del Sacro Cuore},
    addressline={via Branze 45}, 
    city={Brescia},
    postcode={I-25121}, 
    country={Italy}}
    
\affiliation[5]{organization={Politecnico di Torino, Department of Energy},
    addressline={Corso Duca degli Abruzzi 24},
    city={Torino},
    postcode={10129},
    country={Italy}}
    
\affiliation[6]{organization={Dipartimento di Fisica "E. Fermi", Universit\`a di Pisa},
    addressline={Largo B Pontecorvo 3}, 
    city={Pisa},
    postcode={I-56127},
    country={Italy}}
    
\affiliation[7]{organization={ NEST, CNR - Istituto Nanoscienze and Scuola Normale Superiore},
    addressline={piazza San Silvestro 12}, 
    city={Pisa},
    postcode={I-56127},
    country={Italy}}

\affiliation[8]{Institut Universitaire de France (IUF)}

\credit{Data curation, Writing - Original draft preparation}

\cortext[cor1]{First author}
\cortext[cor2]{Principal corresponding author}



\begin{abstract}
Generation of ultra high frequency acoustic waves in water is key to nano resolution sensing, acoustic imaging and theranostics. In this context water immersed carbon nanotubes (CNTs) may act as an ideal optoacoustic source, due to their nanometric radial dimensions, peculiar thermal properties and broad band optical absorption.
The generation mechanism of acoustic waves in water, upon excitation of both a single-wall (SW) and a multi-wall (MW) CNT with laser pulses of temporal width ranging from 5 ns down to ps, is theoretically investigated via a multi-scale approach. We show that, depending on the combination of CNT size and laser pulse duration, the CNT can act as a thermophone or a mechanophone. 
As a thermophone, the CNT acts as a nanoheater for the surrounding water, which, upon thermal expansion, launches the pressure wave. As a mechanophone, the CNT acts as a nanopiston, its thermal expansion directly triggering the pressure wave in water. 
Activation of the mechanophone effect is sought to trigger few nanometers wavelength sound waves in water, matching the CNT acoustic frequencies. 
This is at variance with respect to the commonly addressed case of water-immersed single metallic nano-objects excited with ns laser pulses, where only the thermophone effect significantly contributes. 
The present findings might be of impact in fields ranging from nanoscale non-destructive testing to water dynamics at the meso- to nano-scale.
\end{abstract}



\maketitle

\section{Introduction}
The search for a viable way of generating ultra high frequency acoustic waves in water has been booming recently.
Their generation via photoacoustic techniques promises to play a pivotal role in fields ranging from theranostics \cite{moore2019strategies}, nanoscale acoustic imaging \cite{lee2018efficient,mallidi2009multiwavelength,li2015gold,chen2021gold} to water dynamics at the nanoscale \cite{mante2014probing,chakraborty2017can,sun2020observation,yu2021nanoparticle,daisuke2021interaction,zakhvataev2021existence, fasano2019thermally} and in the hypersonic frequency range \cite{sun2018femtosecond}, below the elastic spectral window accessible to inelastic X-ray, Brillouin inelastic ultraviolet and neutron scattering techniques \cite{sette1995collective,ruocco1999high,bencivenga2007high,ruocco2008history,bencivenga2009temperature}.

\begin{figure*}[]
\centering
\includegraphics[scale=0.17]{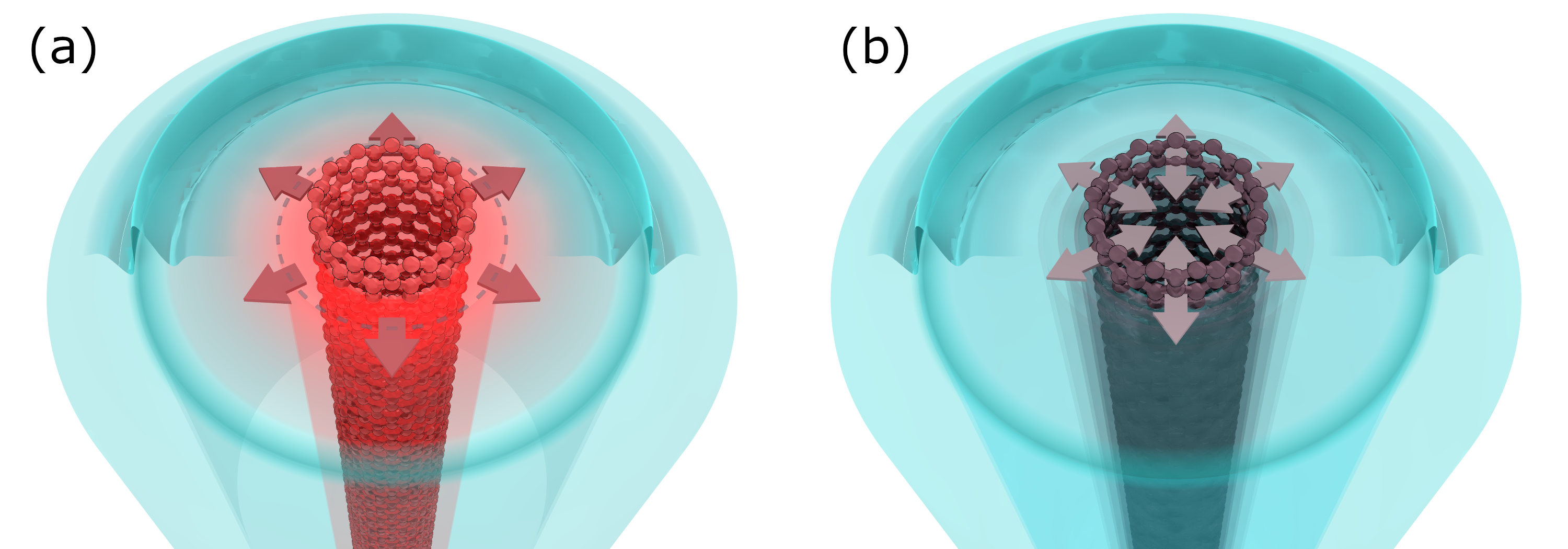}
\caption{Thermophone (a) and mechanophone (b) effects.
The acoustic wave in water is generated in the thermophone case (a) by the thermal expansion of water, the CNT acting as a nanoheater; in the mechanophone case (b) by the thermal expansion of the CNT, the latter acting as a nanopiston.}
\label{fig:Thermophone vs Mechanophone}
\end{figure*}

In a nutshell, the technique is based on heating up an optically absorbing nano-object via a laser pulse. The absorber then launches acoustic waves either by (a) heat transfer to the surrounding water, thus triggering water thermal expansion, or (b) direct thermal expansion of the absorber itself, which acts as a piston for the aqueous medium.
In the former case the transducer nano-object is referred to as a \textit{thermophone} \cite{guiraud2021thermoacoustic,guiraud2019multilayer}, in the latter as a \textit{mechanophone}, see Fig. \ref{fig:Thermophone vs Mechanophone}.

So far, most of the investigations focused on ns laser excitation of metallic nanoparticles, either single \cite{gandolfi2020optical,prost2015photoacoustic,hatef2015analysis,kumar2018simulation,chen2012environment,shahbazi2019photoacoustics,shi2017quantifying,pang2019theoretical} or agglomerates \cite{chen2012environment}, and on carbon nanotubes (CNTs) based composites \cite{lee2018efficient, pramanik2009single,baac2012carbon}, ns laser being the standard for photoacoustic generation in bio-medical applications. Few studies addressed shorter laser pulses on agglomerates of single-wall (SW) CNTs \cite{golubewa2020single}, the role of the nanometric size thus being biased. The implications of the laser pulse duration on photoacoustic generation of a single gold nanosphere was marginally touched by the seminal work of Prost et al. \cite{prost2015photoacoustic}, whose main focus though was on the long pulsed-regime.
In all cases, with the exception of Prost et al. \cite{prost2015photoacoustic}, only the thermophone effect was active. Activation of the mechanophone effect is sought to trigger few nanometers wavelength sound waves in water.

In this context, single water-immersed CNTs could act, in conjunction with short laser pulses in the visible spectrum, as an ideal optoacoustic source, due to their nanometric radial dimensions, high thermal conductivity and broad band optical absorption. The present work theoretically addresses the generation mechanism of acoustic waves in water upon excitation of both a SWCNT and a multi-wall (MW) CNT with laser pulses of temporal width ranging from 5 ns down to ps. 
We demonstrate that, depending on the combination of CNT size and laser pulse duration, the CNT can act as a thermophone or a mechanophone. Activation of the mechanophone effect can trigger, for short enough laser pulses, few nanometers wavelength sound waves in water, matching the CNT acoustic frequencies. This scenario is in contrast with respect to the state of the art, and has potential impact both for nano-acoustic applications and fundamental aspects of water dynamics.

Methodologically, we adopt a multi-physics, multi-scale approach, a strategy that has been shown as very effective in addressing ultrafast opto-thermal \cite{caddeo2017thermal,banfi2010ab} and opto-acoustic \cite{rizzi2021analytical,benetti2017bottom,ronchi2021discrimination} transients in nano to mesoscale systems. In this work, the fact that water infiltrates the CNT or not and the thermal boundary conductance (TBC) between the CNT and water, a key quantity ruling the dynamics, are obtained from atomistic simulations \cite{mohammad2019heat, tascini2017thermal}. The impulsive opto-thermo-mechanical dynamics is then cast in the frame of a continuum model, upon insertion of the geometrical and thermal microscopic parameters, and solved via Finite Element Methods (FEM).

\section{Theoretical model}
The theoretical frame conceived to address the transient photothermal-acoustic response of water-immersed individual CNTs triggered by an ultrafast laser pulse is inherently multiphysics. It involves optics (absorption of the laser pulse by the system), thermics and acoustics. We rely on a macro-physics approach based on continuum equations, upon insertion of the TBC at the CNT/water interface retrieved from dedicated atomistic simulations.

The following assumptions apply: (i) cylindrical symmetry: the CNT height/diameter ratio is large enough to neglect border effects close to the top/bottom of the CNT; (ii) materials thermal and mechanical parameters are temperature independent, the variations being exiguous: their values are taken as the average over the temperature span resulting from simulations; (iii) linear elasticity: the resulting displacement and pressure fields are small enough to justify this assumption; (iv) negligible thermo-acoustic feedback: the thermal dynamics affects the mechanical one, but the opposite effect is negligible; (v) the thermal and mechanical parameters are frequency independent; (vi) non viscous water is assumed: intrinsic dissipation in water is hence not present. This last assumption is an idealization at high frequencies, nevertheless (a) it permits to understand the physics behind the generation mechanism; (b) it provides results based on solid basis, the dissipation mechanism in water in the hypersonic frequency range being yet a debated issue.
\\
\subsection{CNT modeling}\label{subsec:CNTmodelization}
\begin{figure}[h]
\centering
\includegraphics[scale=0.75]{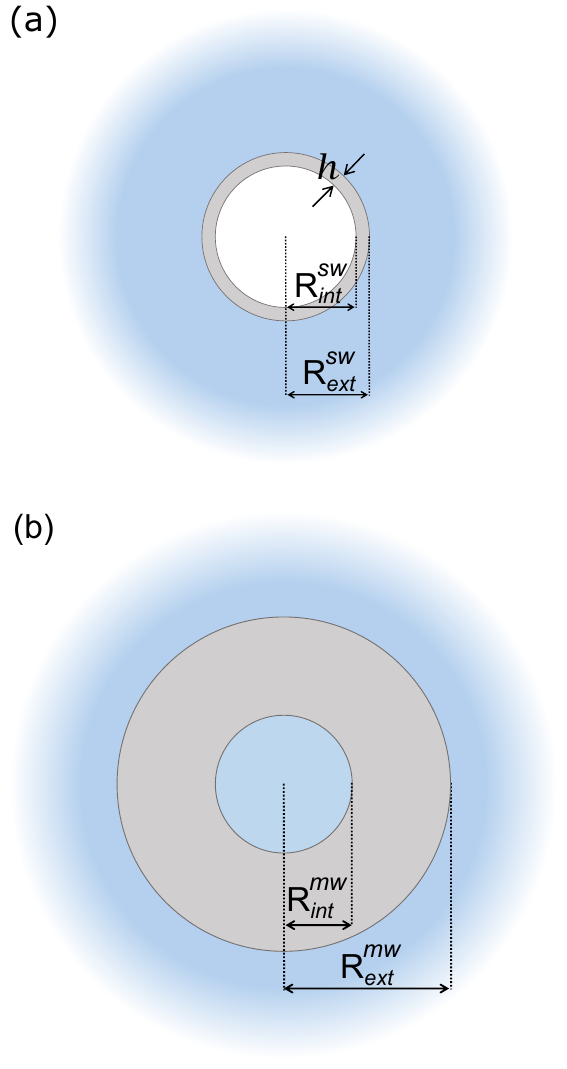}
\caption{Cross-sectional view at constant $z$ of the continuum model for the water-immersed (a) SWCNT: $R_{int}^{sw}$=0.17 nm and $R_{ext}^{sw}=0.51$ nm; (b) MWCNT: $R_{int}^{mw}$=7.3 nm and $R_{ext}^{mw}=15.5$ nm. The MWCNT is water filled, at variance with the SWCNT. The figures are not to scale.}
\label{fig:CNT scheme}
\end{figure}

We address two paradigmatic cases: the case of a single, water-immersed, (5,5) SWCNT and that of a MWCNT composed of N=24 walls. Molecular dynamics (MD) simulations show that water actually infiltrates the MWCNT, whereas the SWCNT remains hollow, its internal radius being too small to accommodate water molecules, see Appendix \ref{MD}.
The time and lengths scales involved in the photoacoustic dynamics are too extended to be addressed in the frame of atomistic simulations, thus requiring a continuum approach.
To this end, the atomistic view of a CNT is here replaced by a continuum cylindrical shell. The thickness $h$ of a single wall is taken as the interlayer distance between graphite’s sheets ($h$ = 0.34 nm), a common choice in CNT modeling \cite{tu2002single, lu1997elastic}. In this way, the (5,5) SWCNT internal and external radii read $R_{int}^{sw}$=0.17 nm and $R_{ext}^{sw}=0.51$ nm.\footnote{Since the nominal center radius for a (5,5) SWNCT is 0.34 nm, its internal and external radii are taken as 0.34 nm $\pm h/2$.} In parellel we take $R_{int}^{mw}$=7.3 nm and $R_{ext}^{mw}=15.5$ nm for the MWCNT. The CNT is assumed infinitely extended along the axial direction, $z$. The external water is accounted for via a coaxial cylindrical shell domain of external radius $R_{wt}$ far exceeding the CNT external radius. A cross-sectional view of the two modelled continuum systems is shown in Fig. \ref{fig:CNT scheme}.
\\
\begin{table*}
  \caption{Material properties used in the continuum model.}
  \label{table:materialproperties}
  \begin{tabular}{llll}
    \hline
    Material property & SWCNT & MWCNT  & Water \\
    \hline
    Radial thermal conductivity [W/m K] & 6 (see text) & 6   & 0.6  \\
    Specific heat [J/kg K] & 650 & 650  & 4.2$\times 10^3$ \\
    Density [kg/m$^3$] & 2.3$\times 10^3$  & 2.3$\times 10^3$ &  1$\times 10^3$  \\
    First Lam\'{e}  coefficient [GPa] & / & / & 2.25  \\
    Second Lam\'{e}  coefficient [GPa] & /  & /  & 0  \\
    Young modulus [GPa] & $1\times10^{3}$ & 30 & /  \\
    Poisson ratio & 0.15  & 0.28  & /  \\
    Linear thermal expansion coefficient [K$^{-1}$] & $3\times10^{-7}$ & $2.5\times10^{-5}$   & $7\times10^{-5}$  \\
    Internal radius [nm] & 0.17  & 7.3  & /  \\
    External radius [nm] & 0.51  & 15.5  & /  \\
    Number of walls & 1  & 24  & /  \\
    \hline
  \end{tabular}
\end{table*}
\subsection{Optics}\label{subsec:Optics}
The dynamics is triggered by a laser pulse impinging on the CNT-water system.
The light intensity is assumed to be spatially homogeneous and gaussian in time:
\begin{equation}\label{}
I(t) = 2\sqrt{\frac{\ln{(2)}}{\pi}}  \frac{\Phi}{\tau}\exp{\Big(-4\ln{(2)}\frac{(t-t_0)^2}{\tau^2}\Big)},
\end{equation}
where $I(t)$ is in W/m$^2$, $\tau$ is the pulse temporal full-width-half-maximum (FWHM), $t_0$ the time at which the pulse intensity is at its maximum (in our case $t_0=0$) and $\Phi$ is the light fluence (J/m$^2$).
In all simulations $\Phi$=5 J/m$^2$, i.e. the total energy carried by the light pulse remains constant, irrespective of the pulse temporal width $\tau$.\\
The laser wavelength is chosen in the water transparency window, 400-700 nm, i.e. the therapeutic spectral window in theranostic applications. The water temperature rise is hence only due to the heat flux from the CNT.\\
The power density (W/m$^3$) absorbed by the CNT reads:
\begin{equation}\label{}
Q_{cnt}(t)=\frac{\sigma_{cnt}^{abs}  I(t)}{ V_{cnt}},
\end{equation}
where $V_{cnt}$ (m$^3$) and $\sigma_{cnt}^{abs}$ (m$^{2}$) denote the CNT volume and absorption cross section, respectively\footnote{Note that both $\sigma_{cnt}^{abs}$ and $V_{cnt}$ are proportional to the CNT length, $Q_{cnt}(t)$ thus being independent on length.}.
The quantity $\sigma_{cnt}^{abs}$ is calculated summing the contributions of all C-atoms in the CNT, assuming they all exhibit the same absorption cross section\footnote{This value corresponds to the one for a flat graphene sheet and is retrieved experimentally and theoretically in CNT for spectrally wide excitation or for large CNT diameters \cite{blancon2013direct, vialla2014universal}.}, $\sigma_{C}^{abs}$=6$\times$10$^{-22}$ m$^{2}$, independently of the wall to which they belong.
\\
$I(t)$, and consequently $Q_{cnt}$, are taken spatially uniform, despite the fact that, in a actual experiment, the laser pulse encounters one side of the CNT first (assuming the pulse propagating perpendicular to the $z$-axis). The high thermal conductivity within the CNT graphene layers, together with the exiguous CNT radial dimension and absorption cross-section, lead to a spatially uniform $Q_{cnt}$ on a time scale inferior to the time-scales involved in (i) heat propagation to water and (ii) the mechanical response.

Summing up, the absorbed power density throughout the system reads:
\begin{equation}\label{}
  Q(t)=\begin{cases}
    Q_{cnt}(t) & \text{in the CNT}\\
    0 & \text{in water}.
  \end{cases}
\end{equation}
\\
\subsection{Thermal dynamics}\label{subsec:Thermal_dynamics}
The thermal dynamics, initiated by the absorbed power density, $Q(t)$, is addressed via the Fourier's law and the continuity equation in cylindrical coordinates, their radial components reading:
\begin{equation}\label{eq:thermics}
\begin{split}
& {q_r}=-k \frac{\partial T}{\partial r}
\\ &\rho c_p \frac{\partial T}{\partial t} = -\frac{1}{r}q_r -\frac{\partial q_r}{\partial r} + Q(t),
\end{split}
\end{equation}
where $T$ is the temperature (K), $q_{r}$ the radial heat flux (W/m$^2$) and the material parameters $k$, $\rho$ and $c_p$ are the radial thermal conductivity (W/(m$\cdot$K)), mass density (kg/m$^3$) and specific heat (J/(kg$\cdot$K)) of the CNT and water in their respective domains.

The radial thermal conductivity of the MWCNT closely matches the out-of-plane thermal conductivity of graphite \cite{che2000thermal} whose value is 6 W/(m$\cdot$K). Atomistically, the concept of radial thermal conductivity does not apply for a SWCNT since it is made of a single carbon layer. On the contrary, modeling the SWCNT as a cylindrical shell of \textit{finite} thickness $h$ requires introducing an artificial radial thermal conductivity in order for Eqs. \ref{eq:thermics} to be meaningful. The value should be taken high enough so as to avoid a temperature gradient across the CNT thickness on the time scales of interest. Given the minute value of $h$, this is achieved even adopting a such small $k$ value as the one for the MWCNT\footnote{We tested values ranging from 6 W/(m$\cdot$K) up to 6 orders of magnitude larger with no difference in the simulations outcome.}. For the CNT specific heat of both SWCNT and MWCNT, we consider 650 J/(kg$\cdot$K) as for graphite \cite{hone2002thermal,pradhan2009specific}. The values of the adopted materials parameters are summarized in Table \ref{table:materialproperties}.

The heat flux at the CNT-water interface, $r$=$R$, is regulated via the following boundary conditions, describing the continuity of the heat flux and the temperature jump:
\begin{equation}\label{eq:thermalboundary}
\begin{split}
& q_r(R-) = q_r(R+)= q_r(R)
\\ & q_r(R)= G  \big[ T(R-)-T(R+) \big],
\end{split}
\end{equation}
where $R-$ and $R+$ refer to the inner and outer side of the interface respectively, and $G$=2.86$\times$10$^{6}$ W/(m$^2$K) is the thermal boundary conductance retrieved from dedicated approach to Equilibrium Molecular Dynamics (EMD) simulations, as detailed in Appendix \ref{MD}. The same value for $G$ is been adopted both for the SWCNT and the MWCNT interfaces with water. For the case of the SWCNT, $R$=$R_{ext}^{sw}$, there being only the external CNT-water interface; whereas a thermal insulating boundary condition, $q_r$=0, is set at the internal CNT-vacuum boundary $r$=$R_{int}^{sw}$, see Fig. \ref{fig:CNT scheme} (a). For the case of the MWCNT instead, $R$=$R_{int}^{mw}$ at the internal, and $R$=$R_{ext}^{mw}$ at the external CNT-water interface, respectively, see Fig. \ref{fig:CNT scheme} (b). In all cases, a thermal insulating boundary condition is enforced at the water domain outer boundary, $q_{r}(R_{wt})=0$.

Before the laser pulse strikes, the system is at the ambient temperature $T_{eq}$=293 K.
\subsection{Mechanical response}\label{subsec:Mechanical_response}
With the spatio-temporal temperature scalar field acting as a source term, we can access the stress fields throughout the entire system and, ultimately, the pressure field in water.
The equations governing the mechanical response read \cite{auld1973acoustic}:
\begin{equation}\label{ElasticityEq}
\begin{split}
& \rho \frac{\partial^2 \textbf{u}}{\partial t^2} =  \boldsymbol{\nabla} \cdot \boldsymbol{\sigma}
\\ & \boldsymbol{\sigma} = \textbf{C} \cdot \big( \boldsymbol{\nabla}_s \textbf{u}-\boldsymbol{\alpha} \Delta T \big),
\end{split}
\end{equation}
where $\textbf{u}$ is the displacement field (m), $\boldsymbol{\sigma}$ the stress tensor (Pa), $\boldsymbol{\nabla}_s \textbf{u}$ the symmetric part of the displacement gradient tensor operator and $\Delta T$ the temperature difference $T-T_{eq}$. The material parameters $\textbf{C}$ and $\boldsymbol{\alpha}$ are the stiffness (Pa) and the linear thermal expansion (1/K) tensors  of the CNT and water in their respective domains.
Anticipating that the thermal expansion can be here taken as a scalar, $\alpha$, and exploiting cylindrical symmetry, Eq. \ref{ElasticityEq} may be limited to its radial and angular components, uncasting the form:
\begin{equation}\label{eq:Sigma}
\begin{split}
& \rho \frac{\partial v_{r}}{\partial t} = \frac{\partial \sigma_{rr}}{\partial r}+\frac{1}{r}(\sigma_{rr}-\sigma_{\phi \phi})
\\ & \frac{\partial \sigma_{rr}}{\partial t} = c_{11} \frac{\partial v_r}{\partial r} + c_{12} \frac{v_r}{r} - \alpha  \big(c_{11}+c_{12}+c_{13} \big)\frac{\partial T}{\partial t}
\\ & \frac{\partial \sigma_{\phi \phi}}{\partial t} = c_{12} \frac{\partial v_r}{\partial r} + c_{11} \frac{v_r}{r} - \alpha  \big(c_{11}+c_{12}+c_{13} \big) \frac{\partial T}{\partial t}
\end{split}
\end{equation}
where $v_r=\partial u_{r}/\partial t$ and $c_{ij}$ ($i,j=1,2,3$) are $\textbf{C}$ components in Voigt notation\footnote{The equation for $\sigma_{zz}$ is not reported since we are only interested in the radial direction and $\sigma_{zz}$ does not appear in the equations for $\sigma_{rr}$ or $v_{r}$.}. 

At the CNT-water interface the continuity of $v_r$ and of $\sigma_{rr}$ applies:
\begin{equation}\label{}
\begin{split}
& v_r(R-) = v_r(R+)
\\ & \sigma_{rr}(R-) = \sigma_{rr}(R+).
\end{split}
\end{equation}
Specifically, for the case of the SWCNT, $R$=$R_{ext}^{sw}$, there being only the external CNT-water interface, whereas a stress-free boundary condition is set at the internal CNT-vacuum boundary, $\sigma_{rr}(R_{int}^{sw})$=0, see Fig. \ref{fig:CNT scheme} (a). For the case of the MWCNT instead, $R$=$R_{int}^{mw}$ at the internal, and $R$=$R_{ext}^{mw}$ at the external CNT-water interface, respectively, see Fig. \ref{fig:CNT scheme} (b). In all cases, a stress-free boundary condition is enforced at the water domain outer boundary, $\sigma_{rr}(R_{wt})=0$, since the water outer boundary is far enough that it is never reached by the sound wave in our simulation time. Before the pulse strikes, the system is at rest and stress free, the initial conditions being $u_r=0$, $v_r=0$ and  $\sigma_{rr}=0$.
A key, yet unsolved issue, concerns the elastic constants to be adopted for the CNTs. Developing a model for CNTs elasticity is not straightforward. CNTs are built of an highly anisotropic material, graphite, which is rolled up and even radially deformed to accomodate for the correct matching of the single atomic cylinders. For these reasons, the literature lacks of an ultimate and unanimous model for the $c_{ij}$ coefficients of CNTs \cite{lu1997elastic,velasco2009vibrations,zaeri2015elastic}.\\
In this work, we proceeded as follows. We modeled the CNT as a hollow cylindrical shell made out of a \textit{homogeneous} and \textit{isotropic} material with Young modulus $E_Y$ and Poisson ratio $\nu$.
In particular, for the SWCNT case, we developed a simple model allowing to calculate $E_Y$ and $\nu$ starting from the known 2D elastic properties of graphene. Notably, the use of the calculated values allows, in a continuum mechanics model, to correctly predict the experimentally observed frequency of the radial breathing modes reported in the literature \cite{maultzsch2005radial}. For the more complex MWCNT case, instead, we rely on experimental values found in the literature \cite{palaci2005radial}. Both these methods are detailed in Appendix \ref{CNT Mechanics}. The actual values of $E_Y$ and $\nu$ are reported in Table \ref{table:materialproperties}. We note that this approach can lead to imprecise estimates for vibrational modes other than the radial ones (as for instance a lateral bending). The model though is precise enough for our purpose, correctly describing elasticity in the radial direction and retaining the symmetries of the problem. Furthermore, its simplicity allows for a straightforward physical insight.\\
Once obtained $E_Y$ and $\nu$, we calculate the $c_{ij}$ coefficients for the CNT pertinent to our problem:
\begin{equation}\label{elastic conversion}
\begin{split}
& c_{11}= \frac{E_Y}{1+\nu} \left(1+\frac{\nu}{1-2\nu}\right)
\\ &  c_{12}= \frac{E_Y}{1+\nu} \left(\frac{\nu}{1-2\nu}\right)
\\ &  c_{13}=c_{12},
\end{split}
\end{equation}
whereas, for the water domain, the elastic coefficients are expressed in terms of water's first Lam\'{e} parameter $\lambda$ (Pa), $c_{11}$=$c_{12}$=$c_{13}$=$\lambda$.
These values, together with water thermal expansion, SWCNT's \cite{li2005axial} and MWCNT's\footnote{For the MWCNT, $\alpha$ closely matches the out-of-plan graphite's one \cite{bandow1997radial, maniwa2001multiwalled, marsden2018modelling}.} radial thermal expansion coefficients are reported in Table \ref{table:materialproperties}.
%

We are ultimately interested in the pressure wave launched in the water domain outside the CNT, which is obtained, after numerical solution of Eq. \ref{eq:Sigma}, as $p=-\sigma_{rr}=-\sigma_{\phi\phi}$.\\
In order to discriminate between the two contributions to the pressure wave, we selectively activate the sole thermophone or mechanophone effect setting to zero the thermal expansion coefficient $\alpha$ in the CNT or external water domain, respectively. Indeed, the total pressure field is obtained taking into account both values of the thermal expansion throughout the entire system.
\section{Results and discussion}
\subsection{Thermal dynamics}
\begin{figure*}[t]
\centering
\includegraphics[scale=0.85]{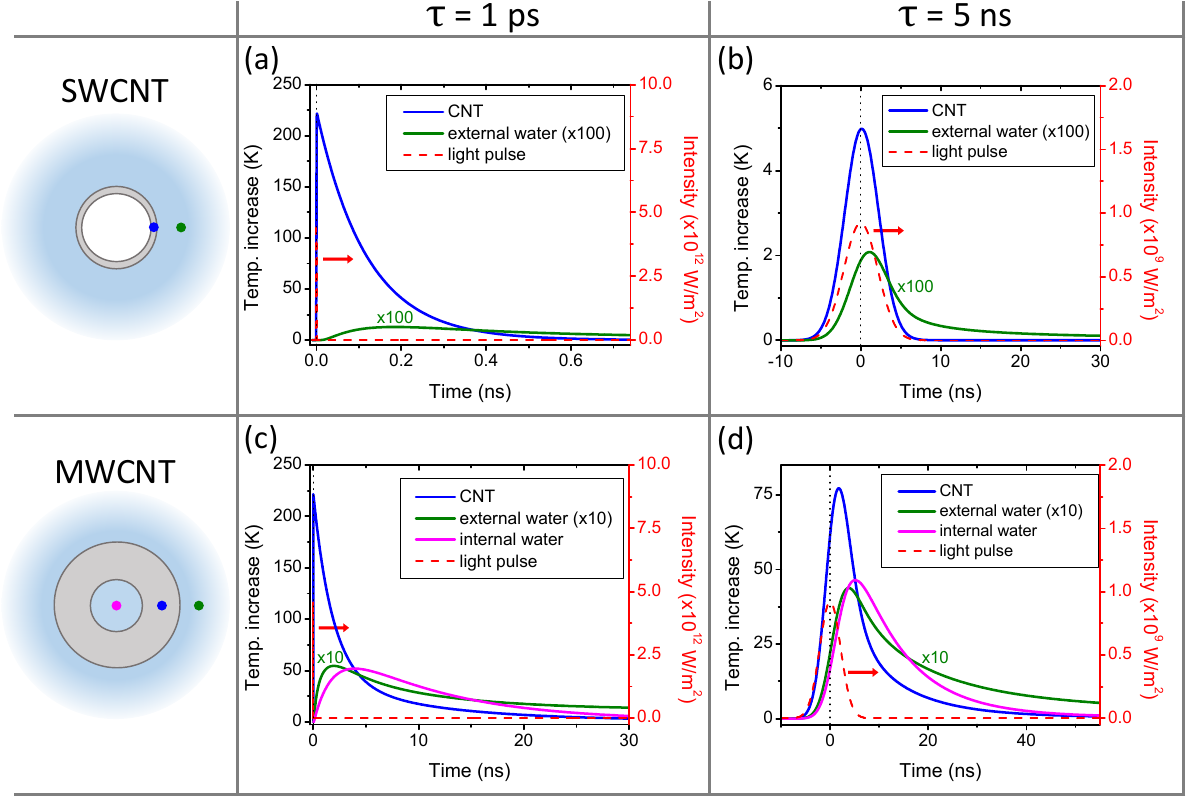}
\caption{Time evolution of the temperature increase in the different CNTs/light pulse duration cases: SW/ps (a), SW/ns (b), MW/ps (c) and MW/ns (d).
Left axis: CNT temperature increase (blue line) in the middle of the CNT radius (blue dot), external water temperature increase (green line) $5$ nm away from the external CNT/water interface (green dot) and, only for the MWCNT case, the internal water temperature increase (magenta line) at the radial coordinate $r=0$ (magenta dot). The external water temperature increase curves are multiplied by a factor 100 for the SWCNT case (a,b) and 10 for the MWCNT case (c, d) to ease readability. Right-axis: light pulse intensity (red dashed line) as a function of time. Maximum intensity in panels (a) and (c) is $\approx 5 \times  10^{12}$ W/m$^2$. The vertical black dotted line labels the instant at which the peak pulse intensity occurs. The time scale differs in each panel.}
\label{fig:Temp_Delta_matrix}
\end{figure*}
We now report simulations results for the thermal dynamics triggered by a laser pulse of duration $\tau$=1 ps (first column of Fig. \ref{fig:Temp_Delta_matrix}) and 5 ns (second column) in the SWCNT (first row) and MWCNT (second row) systems.
%
Each panel reports, on the left axis, the temperature increase $\Delta T$=$T(t)$-$T_{eq}$ calculated at the mid point of the CNT thickness (blue line and dot), at a point of the external water positioned $5$ nm away from the external CNT/water interface (green line and dot) and, only for the case of the MW, in the internal water at $r=0$ (magenta line and dot). The right axis pertains to the light pulse intensity time-dependence (red dashed line).

In general, the CNT temperature increases due to absorption of the laser pulse. The CNT cools down delivering heat to the proximal water, which increases its temperature. Finally, the CNT and the proximal water relax to their initial temperature via heat diffusion to the water bulk.\\
In the \textit{SW/ps pulse case} (Fig. \ref{fig:Temp_Delta_matrix} a), the relative CNT temperature rises up to $\Delta T \approx 225$ K on the time scale of the laser pulse duration $\tau$=1 ps. 
The CNT temperature then relaxes exponentially with a time constant $\tau_{th}$ = $\rho c_{p}R^{sw}_{ext}/2G \approx$ 130 ps\footnote{The formula given for $\tau_{th}$ is actually an approximation valid for $R^{sw}_{int} \ll R^{sw}_{ext}$. The real case formula reads $\rho c_{p}R^{sw}_{ext}(1-(R^{sw}_{int}/R^{sw}_{ext})^2)/2G$, which its' the ratio of the external SCNT surface (the only surface exchanging heat) to the CNT volume. Its validity  is assured by a Biot number $Bi=G R^{sw}_{ext}(1-(R^{sw}_{int}/R^{sw}_{ext})^2)/2k \ll 1$ and an essentially isothermal water (as compared to the CNT temperature excursion), and indeed matches the value of $\tau_{th}$ obtained fitting the numerical curve. For sake of completeness, we pinpoint that, substituting $R^{sw}_{int}=R^{sw}_{ext}-h$ in the real case formula, and expanding to first order in $h/R^{sw}_{ext}$ for $h/R^{sw}_{ext} \ll 1$, one obtains $\rho C_{p}h/G$. In the present case though $h/R^{sw}_{ext}\sim 0.7$, the latter approximation hence being flaw.}, releasing heat to water, whose $\Delta T$ reaches a maximum of a fraction of a degree.
The message to retain here is that the CNT is heated up on a time scale $\tau$ and remains hot and thermally decoupled from water on a much longer time scale, $\tau_{th}$, ruled by thermal boundary conductance. Under these circumstances, arising for $\tau\ll\tau_{th}$, one may intuitively expect the mechanophone effect to play a role, a statement to be formally proven upon calculating the mechanical response.\\
In the \textit{SW/ns pulse case} (Fig. \ref{fig:Temp_Delta_matrix} b) the situation is the opposite: $\tau \gg \tau_{th}$. The light pulse temporal width, now $\tau$=5 ns, exceeds by far $\tau_{th}$=130 ps, implying that the heat dissipation rate from the CNT to water matches, on the ns time-scale, the energy absorption rate from the CNT. This is evident noticing that the CNT $\Delta T$ temporal evolution precisely matches the light pulse temporal one, $I(t)$, releasing heat fast enough to the external water, and ending with $\Delta T \approx 0$ at the instant in which the light pulse is terminated, i.e. $t \approx 5$ ns. The water $\Delta T$ temporal evolution, at a point 5 nm out of the CNT/water boundary, matches almost as well the CNT's one (delay between the two curves $\approx$ 1 ns). Otherwise stated, the scenario is that of steady state heat transfer under continuum illumination with a slowly (with respect to the thermal transient time from the CNT to water, i.e. $\tau_{th}$) modulated heat source (the modulation being the Gaussian pulse profile lasting $\sim\tau$). This scenario is the one commonly encountered in nanofluids under irradiation with 5 ns pulses \cite{gandolfi2020optical}, thus we expect the thermophone effect being the dominating photoacoustic launching mechanism.\\
\textit{MW/ps case} (Fig. \ref{fig:Temp_Delta_matrix} c). Switching to the MWCNT, two new aspects set in: an increased CNT volume and the presence of the internal water. Both these aspects imply an overall longer thermal relaxation time of the CNT $\Delta T$, characterized by a non-exponential temperature decay. We still keep for good the concept of $\tau_{th}$ as the time scale over which the CNT relaxes heat to the external water, even though, in this case, due to the more complex thermalization dynamics, $\tau_{th}$ is not anymore an exponential decay but we rather define it as the time it takes for the CNT $\Delta T$ to fall to 1/e of its maximum value, resulting in $\tau_{th}\approx 5$ ns. The latter definition, although somewhat arbitrary, provides a sound rule-of-thumb estimate for the CNT thermal relaxation time to the external water. The reasons for the increased $\tau_{th}$ in the MWCNT case stand in the augmented thermal mass of the CNT and the added contribution of the internal water thermal inertia.\\
We now dwell on the details of the thermal dynamics to substantiate the above-mentioned concepts. Always with reference to Fig. \ref{fig:Temp_Delta_matrix} c, the CNT has two heat thermal pathways in parallel, namely to the internal and to the external water. 
The CNT initially cools, delivering heat to both water domains. The internal water $\Delta T$ becomes though rapidly higher with respect to that of the external water and, after $t \approx$ 5 ns, attains the same value as that of the CNT.
Starting from this instant, the internal heat flux is reversed, flowing from the internal water domain to the CNT. This slows down the cooling rate of the CNT, which, while dissipating heat towards the external water, absorbs heat form the internal one. This explains why, starting from 5 ns, the internal water temperature overshoots the CNT's one.
For our purposes, the important message is that, also for the MW/ps case, $\tau\ll\tau_{th}$, thus suggesting that the mechanophone should be also effective.\\
In the \textit{MW/ns case} (Fig. \ref{fig:Temp_Delta_matrix} d) the laser pulse duration $\tau=5$ ns $\approx$ $\tau_{th}$. This situation is a cross-over between the SW/ns and the MW/ps. The mechanophone effect could potentially be active, but to a lesser extent with respect to the MW/ps, where $\tau\ll\tau_{th}$. Consistently with the above discussion, and contrary to the SW/ns case, the situation is not, strictly speaking, that of steady state heat flow, as can be appreciated noting that the CNT $\Delta T$ profile does not match that of $I(t)$.

Summing up, the solutions to the thermal problem, first, provide the source terms for the mechanics and, second, suggest that the mechanophone effect should not be expected in the SW/ns case but could be active in the SW/ps and MW/ps cases, and perhaps, in the MW/ns one.
\subsection{Mechanical response}
\begin{figure*}[t!]
\centering
\includegraphics[scale=0.85]{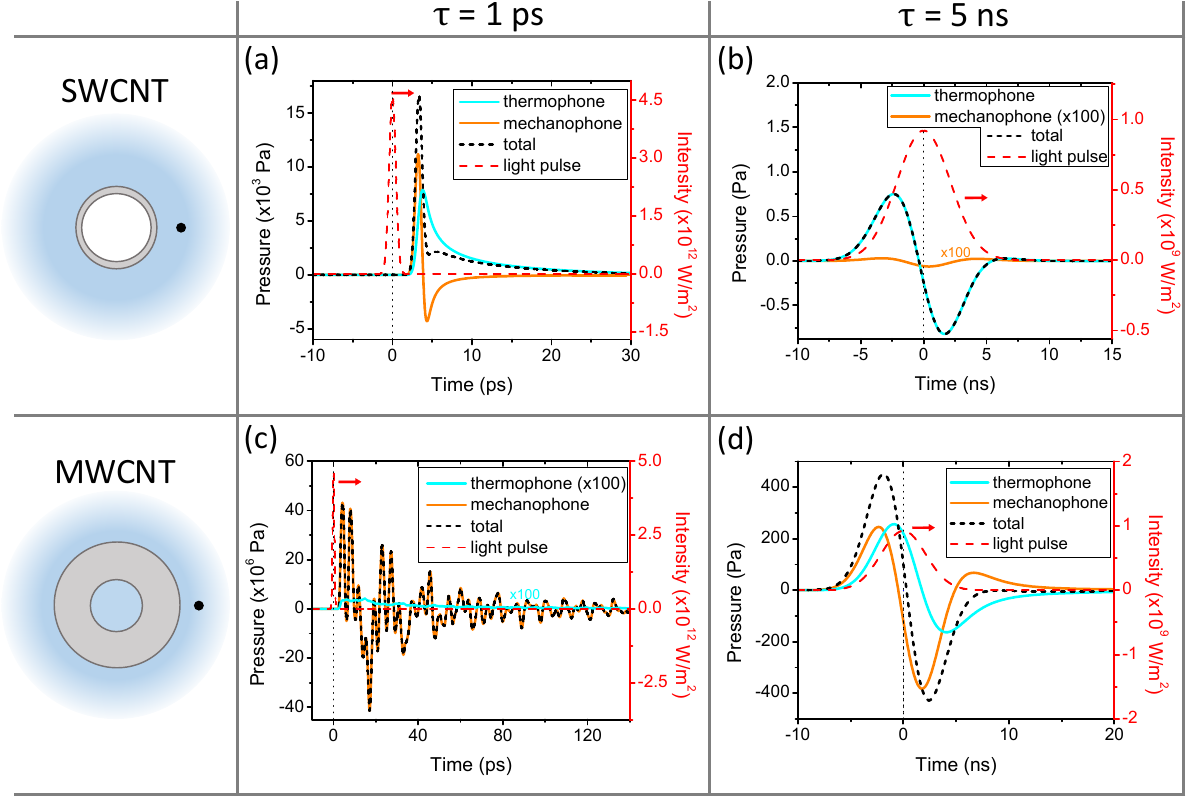}
\caption{Time evolution of the pressure field for the different CNTs/light pulse cases: SW/ps (a), SW/ns (b), MW/ps (c) and MW/ns (d). The pressure is calculated at a point in the external water $5$ nm from the CNT/water interface (black dot). Left axis: thermophone pressure contribution (cyan blue line), mechanophone pressure contribution (orange line) and total pressure (black dotted line). The pressure curves are multiplied by a factor 100 for the mechanophone contribution in the SW/ns case (panel b) and for the thermophone contribution in the MW/ps case (panel c) to ease readability. Right-axis: light pulse intensity (red dashed line) as a function of time. The vertical black dotted line labels the instant at which the peak pulse intensity occurs. The time and pressure scale differs in each panel.}
\label{fig:Matrix_Pressure_vs_Time}
\end{figure*}
We now report simulations results for the mechanical response triggered by a laser pulse duration $\tau$=1 ps (first column of Fig. \ref{fig:Matrix_Pressure_vs_Time}) and 5 ns (second column) in the SWCNT (first row) and MWCNT (second row) systems, the spatio-temporal temperature profile being the source term for the acoustic problem.
Each panel reports, on the left axis, the pressure field due solely to the thermophone (cyan line), solely to the mechanophone effect (orange line) and the total pressure (black dotted line) calculated at a point in the external water $5$ nm away from the CNT/water interface (black dot). The right axis pertains to the light pulse intensity time-dependence (red dashed line), with $t=0$ being the instant at which the peak pulse intensity occurs (vertical dashed line).

The first striking feature, that stems out from a bird-eye view analysis of the four panels, is that the mechanophone effect indeed contributes significantly to the total pressure in all cases except the SW/ns one. Specifically, comparing the peak pressure 5 nm out of the CNT/water interface, the mechanophone and thermophone equally contribute in both the SW/ps (Fig. \ref{fig:Matrix_Pressure_vs_Time} a) and MW/ns cases (Fig. \ref{fig:Matrix_Pressure_vs_Time} d), whereas, for the MW/ps case the mechanophone is the only contributing term (Fig. \ref{fig:Matrix_Pressure_vs_Time} c).\\
This analysis, although providing a vivid representation of the effect, does not suffice to quantitatively address the relative contributions of the two effects. 
\begin{figure}[]
\centering
\includegraphics[scale=0.87]{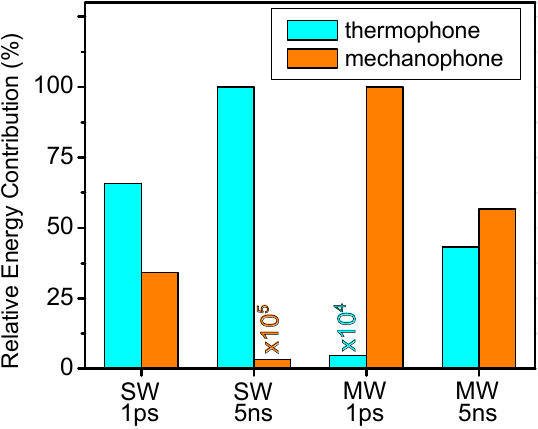}
\caption{Relative energy contribution, in percentage, of the thermophone (cyan bar) vs mechanophone (orange bar) effects calculated from the acoustic Poynting vector for the different cases SW/ps, SW/ns, MW/ps and MW/ns. For the sake of visualization, in the SW/ns case the mechanophone contribution is multiplied by $10^5$, while in the MW/ps case the thermophone contribution is multiplied by $10^4$.}
\label{fig:Histogram Poyting Energy}
\end{figure}
For a quantitative comparison of the two effects we rely on energy considerations. We first calculate, for each size and pulse duration case and accounting for the thermophone and mechanophone effects separately, the acoustic Poynting vector \cite{auld1973acoustic} (W/m$^2$):
\begin{equation}\label{}
\textbf{P} = -v_r \, \sigma_{rr} \ \boldsymbol{\hat{r}},
\end{equation}
where $\boldsymbol{\hat{r}}$ is the radial versor. 
We then evaluate the total mechanical energy transported by the acoustic wave in the external water, calculating the flux of $\textbf{P}$ across a generic cylindrical surface, $S$\footnote{The surface $S(r)$ has been taken for big enough $r$ values so as for the energy $U$ to attain an asymptotic value. This procedure allows accounting for the volume of the water domain actually contributing to the pressure pulse.}, coaxial to and encapsulating the CNT, and integrating the flux in time:
\begin{equation}\label{}
U=\int_t \oint_S \textbf{P} \cdot \boldsymbol{\hat{n}}\ \,dS dt,
\end{equation}
where $\boldsymbol{\hat{n}}$ is the outgoing unit vector perpendicular to the surface $S$.\\
The comparison between the relative thermophone and mechanophone energy contributions for the different CNT size and pulse duration cases is shown in percentage in the histogram of Fig. \ref{fig:Histogram Poyting Energy}. 
For the SW/ps case, the thermophone contribution is comparable to the mechanophone one, exceeding the latter by 60$\%$ (despite the fact that, in Fig. \ref{fig:Matrix_Pressure_vs_Time} a, the thermophone pressure peak is lower than the mechanophone one). 
For the SW/ns case the thermophone dominates, while for the MW/ps case only the mechanophone contributes. 
For the MW/ns case the mechanophone contribution is slightly greater than the thermophone's.

The histogram proves, on a quantitative basis, what was suggested by the solution of the thermal problem. Specifically, the mechanophone effect sets in, on an equal footing with the thermophone's or even as the prevailing contribution, when the pulse duration is comparable or smaller than $\tau_{th}$. 
We pinpoint that the latter condition is not satisfied for metallic nanosystems under 5 ns laser pulse width excitation, as commonly addressed in the literature, where only the thermophone effect plays a role \cite{gandolfi2020optical, chen2012environment, shahbazi2019photoacoustics}.
On the material side, contributing to this difference is the fact that the TBC at the CNT/water interface is about one order of magnitude lower than for the case of typical metal/water interfaces \cite{chen2022interfacial, merabia2009heat, ge2004aupd}. This increases $\tau_{th}$ for the case of the CNT. Otherwise stated, on the one hand, a lower TBC value reduces the water peak temperature increase and thermal expansion, resulting in a smaller thermophone effect; on the other hand, it brings the CNT to a higher temperature and greater thermal expansion, enhancing the role of the mechanophone effect in the acoustic wave generation.
The mechanophone contribution thus emerges as a relevant launching mechanism.\\

\begin{figure}[]
\centering
\includegraphics[scale=0.85]{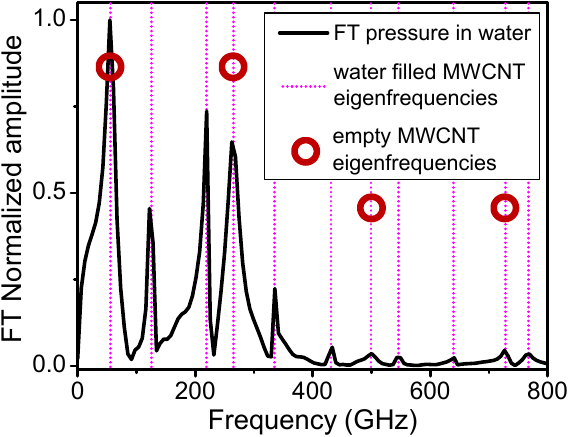}
\caption{Normalized Fourier transform amplitude of the total pressure time evolution curve in water for the MW/ps case (black line).  Eigenvalues of the radial displacement field for the water-filled MWCNT (vertical dotted magenta lines) and for the empty MWCNT (empty dots).}
\label{fig:FT__Normalized_Pressure_MW_1ps_inkscape}
\end{figure}

We now focus on the shape of the pressure pulse in time and on the possibility, in the MW/ps case, to launch hypersonic frequency - few nanometers wavelength acoustic waves in water, matching the CNT mechanical eigenmodes frequencies.\\
In the two \textit{ns cases} (panels b and d of Fig. \ref{fig:Matrix_Pressure_vs_Time}), the total pressure displays the typical profile expected for acoustic waves generated by liquid-embedded nano-systems excited by ns laser pulses \cite{gandolfi2020optical, prost2015photoacoustic,pang2019theoretical, calasso2001photoacoustic}, where a positive peak is followed in time by a negative one, the pressure profile being proportional to the time derivative of the laser pulse intensity \cite{calasso2001photoacoustic}. 
Things are quite different for the case of ps pulse excitation.\\
In the \textit{SW/ps case} (Fig. \ref{fig:Matrix_Pressure_vs_Time} a), the total pressure takes the shape of a sharp positive peak in time, followed, on a longer time scale, by a shallow pressure trailing edge.
Whereas the main contribution to the sharp peak pulse comes from the mechanophone (orange curve), the trailing-edge is mainly contributed by the thermophone effect (cyan curve). After the sharp acoustic pulse is launched, mainly by the CNT expansion, the external water is still rising its temperature, layer by layer as heat propagates radially, resulting in a long lasting pressure tail as a cumulative effect.\\
In the \textit{MW/ps case} (Fig. \ref{fig:Matrix_Pressure_vs_Time} c), water pressure shows damped oscillations in time at different frequencies. The pressure is solely contributed by the mechanophone effect, suggesting that oscillations are related to the eigenmodes of the water-filled MWCNT. This is proven in Fig. \ref{fig:FT__Normalized_Pressure_MW_1ps_inkscape}, where we plot the amplitude, normalized to its maximum, of the Fourier transform (FT) of the total pressure curve for the MW/ps case (black curve). The FT is peaked at the eigenvalues of the displacement field for the water-filled MWCNT system (vertical dotted magenta lines) calculated following Ref. \cite{velasco2009vibrations}. We pinpoint that the infiltrated water does play a role in determining the launched frequencies, the radial displacement eigenvalues of the non-infiltrated MWCNT being a subset of the water-filled MWCNT ones (magenta dots). 
%
These eigenmodes are excited since their frequencies fall within the laser pulse bandwidth (BW) $\approx \tau^{-1}=1$ THz. This is not the case for the MW/ns case, where the laser pulse BW $\approx0.2$ GHz falls short of the fundamental radial eigenfrequency, nor for the SWCNT cases, the SWCNT breathing mode occurring at 10 THz.

\begin{figure}[t!]
\centering
\includegraphics[scale=0.8]{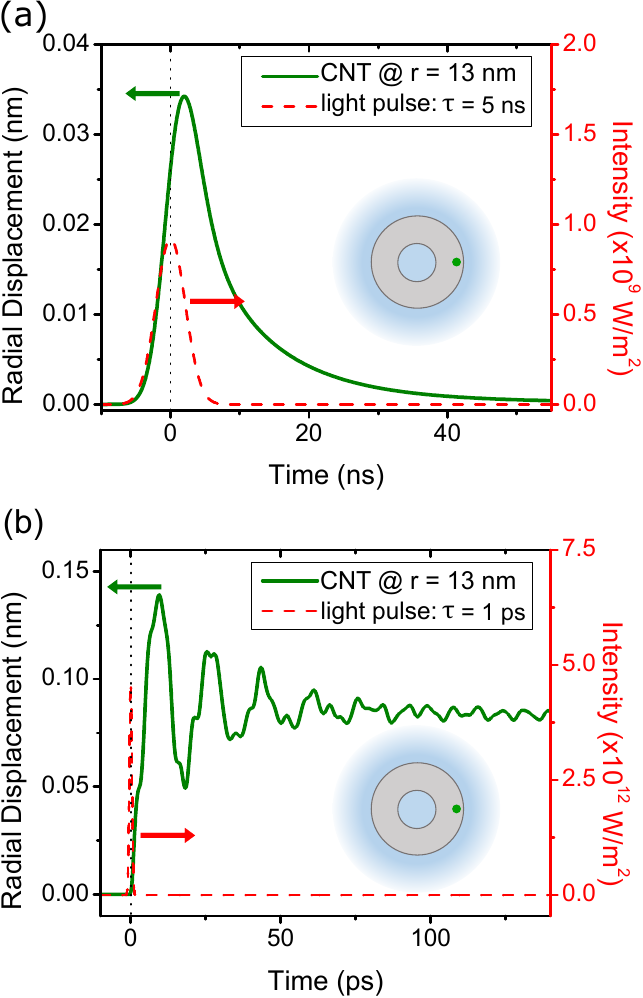}
\caption{Time evolution of the radial displacement (green line, left axis) calculated at $r=13$ nm (green dot in figure inset) for (a) the MW/ns and (b) MW/ps cases. Laser pulse intensity time-dependence (red dashed line, right axis) for the respective cases. Mind the different time scale in each graph.}
\label{fig:Displacement_vs_Time_matrix}
\end{figure}

The appearance of the water-filled MWCNT oscillations and their relation to the laser pulse duration are well visualized in Fig. \ref{fig:Displacement_vs_Time_matrix}, reporting the time evolution of the radial displacement (green line, left axis) calculated at $r$=13 nm, i.e. a point inside the MWCNT wall and close to its external interface with water (green dot in the inset sketch), for the MW/ns case (panel a) and for the MW/ps case (panel b). Mind that the time-scales are different in the two graphs for the sake of visualization.
In the ns case, the radial displacement increases, due to thermal expansion, as long as the CNT accumulates energy, and then returns to zero on the tens of ns time scale while the CNT transfers heat to the surrounding water. In the ps case, oscillations of the CNT vibrational modes are activated by the extended laser pulse BW and superposed on top of the slowly decreasing displacement background. The oscillations are damped via acoustic radiation to the surrounding water on a 100 ps time scale, whereas the average CNT radius relaxes with temperature on a 5 ns time scale. The latter relaxation is not appreciable on the present scale but may be understood from the temperature dynamics of Fig. \ref{fig:Temp_Delta_matrix} c.
\begin{figure*}[h!]
\centering
\includegraphics[scale=0.85]{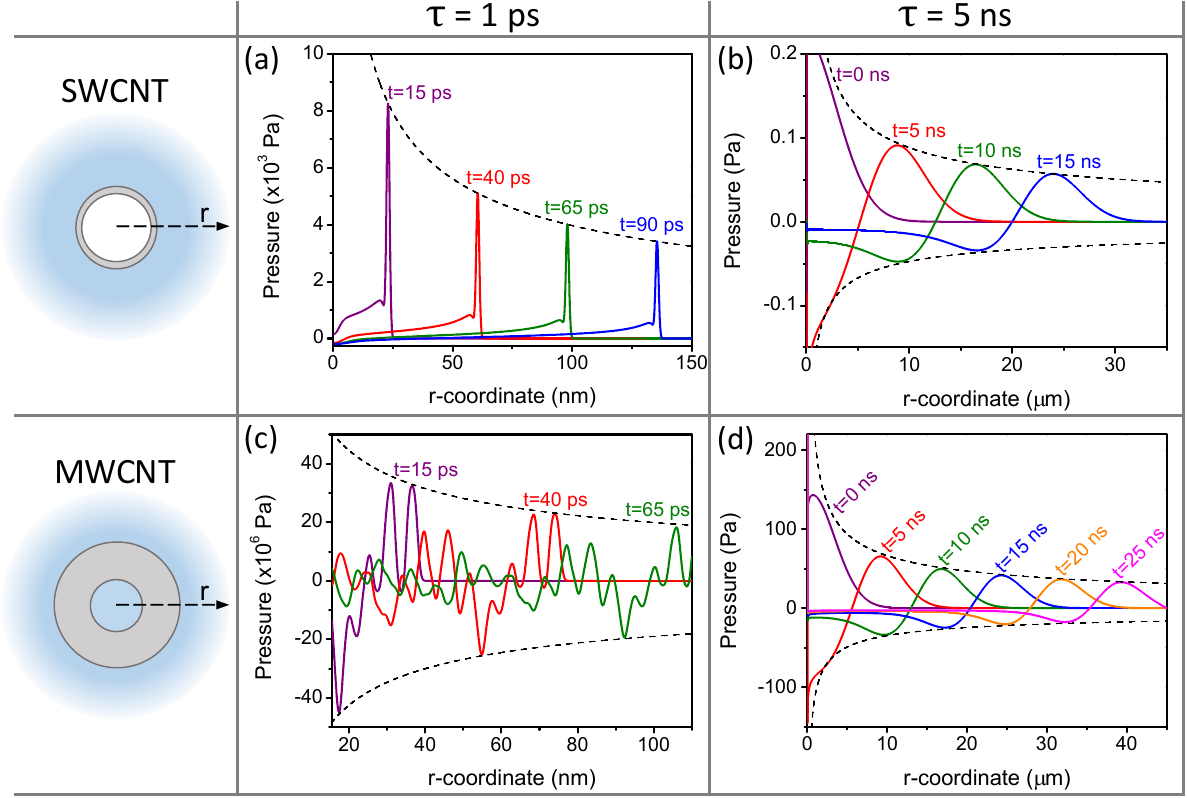}
\caption{Pressure along the radial coordinate in the external water for the different cases: SW/ps (panel a), SW/ns (panel b), MW/ps (panel c) and MW/ns (d). Different colors show different instants in time. The $\propto 1/\sqrt{r}$ envelope to the peaks maxima and minima is shown as a dashed black line for each panel.
}
\label{fig:Matrix_Pressure_radial}
\end{figure*}

For the sake of completeness, we address in Fig. \ref{fig:Matrix_Pressure_radial} the spatio-temporal dynamics. The graphs show the total pressure along the radial coordinate $r$, calculated at different time instants, for all the CNT sizes/light pulse cases. The pressure is shown in the external water domain: for the SW/ps (panel a), SW/ns (panel b), and MW/ns (panel d) the CNT radius is negligibly small compared to the plotted radial range and cannot be appreciated, while for the MW/ps case (panel c) the CNT radius is comparable to the plotted radial range, the $r$ axis hence being reported for $r>$15.5 nm. The envelopes to the highest and lowest peaks (black dotted lines) scale as $1/\sqrt{r}$, which is the expected \textit{asymptotic} behaviour for a cylindrical pressure wave to conserve energy.

For the SW/ns (panel b) and MW/ns (panel d) cases, the typical acoustic wave-like profile, with a positive peak followed by a a negative one, is also retrieved in space.
Differently, for the SW/ps case (panel a), a positive pressure spike clearly stands out, followed by a pressure bulge trailing behind (at smaller $r$ coordinates with respect to the pressure spike). As time evolves, this bulge spreads out and, for radial coordinates closer to the CNT, eventually becomes negative (see Fig. \ref{fig:Matrix_Pressure_radial} at very small delays). This pressure bulge is continuously fed by the thermophone effect. It arises as the cumulative effect of a reduced thermal expansion in proximal water shells that are cooling down, and an increased thermal expansion in water shells that are being reached by the propagation of heat. Although thermal diffusion continuously contributes to the pressure, we recognize the pressure spike, mostly contributed by the mechanophone effect, as the pressure pulse of interest for photacoustics applications.
For the MW/ps case, panel c yields a vivid representation of the short spatial wavelengths involved in the oscillating pressure generated by the water-filled-MWCNT system, here acting as a mechanophone.

At the light of the results, a few \textit{a posteriori} remarks are due concerning some of the adopted assumptions at the basis of our modeling. We adopted cylindrical symmetry. For instance, this approximation is not suitable to model situations where diffraction effects of the acoustic wave packet in water, due to the CNT finite length, become relevant. This occurs when considering the pressure wave front in directions characterized by an high enough polar angle, the $z$-axis being coincident with the CNT longitudinal axis. The materials parameters have been here assumed temperature independent. This aspect might be \textit{a priori} quite critical for the case of water expansion which, contrary to the CNT, is affected by a strong temperature dependence. For high enough water temperature raise, non linear effects are important and might significantly change the water pressure pulse profile \cite{calasso2001photoacoustic}. For the present case, our calculations show that the water temperature increase is at most of the order of few kelvin, see Fig. \ref{fig:Temp_Delta_matrix}, thus permitting us to disregard the water expansion temperature dependence \cite{gandolfi2020optical}. Nevertheless, for the case of higher water temperature excursions (as might arise adopting higher intensity pump pulses or highly absorbing materials) caution should be paid to this issue. As previously explained, in the present work we did not account for intrinsic damping of acoustic waves in water (inviscid water), a yet open issue in the present frequency range. Generally speaking, intrinsic damping is expected to increase with frequency, hence acting as a low pass filter, with minor effects for the case of pressure waves excited with long laser pulses but, in the present case, possibly hampering water oscillations frequencies in excess of the MWCNT fundamental mode at $\approx$ 50 GHz when exciting with ps pulses. For the sake of disentangling the launching mechanisms when exciting with pulses of different temporal width, which is the primary goal of this work, we therefore avoided adding an additional effect, i.e. viscosity, which is yet unclear in the hundreds of GHz frequency range \cite{chakraborty2017can,yu2021nanoparticle}.

As previously explained, in the present work we did not account for intrinsic damping of acoustic waves in water (inviscid water), a yet open issue in the present frequency range \cite{yu2021nanoparticle,chakraborty2017can}.
Intrinsic damping is expected to increase with frequency, hence acting as a low pass filter, with minor effect for the case of pressure waves excited with long laser pulses but a possibly consistent one when exciting with ps pulses.
For the sake of disentangling the launching mechanisms when exciting with pulses of different temporal width, which is the primary goal of this work, we therefore avoided adding an additional effect, i.e. viscosity, mostly affecting the case of short laser pulses and, furthermore, yet unclear in the hundreds of GHz frequency range.
That said, we here give a rough, tentative estimate of how viscosity might affect the pressure wave in water reported in Figure \ref{fig:Matrix_Pressure_radial}. We estimate the acoustic penetration depth, $\mathcal{L}$ (m), for the pressure wave, following Ref. \cite{guiraud2019multilayer}: $\mathcal{L}$ $\approx$ $\nu^{-2}$, it is not valid for viscoelastic fluids (i.e. for frequencies exceeding the inverse of water molecules relaxation time) and it depends upon the viscous water parameters, here taken for the static case. As an indication, the estimate yields $\mathcal{L}\approx$ 100 $\mu$m, 70 nm, 20 nm and 5 nm for $\nu$=1 GHz, 50 GHz, 100 GHz and 200 GHz, respectively. These numbers suggest that intrinsic damping will not be severe for 5 ns laser pulses excitation, but, for instance, in the MWCNT/ps case, will hamper water oscillations frequencies in excess of the MWCNT fundamental mode at $\approx$ 50 GHz.

We estimate the acoustic penetration depth (m) as \cite{guiraud2019multilayer}:
\begin{equation}\label{viscous damping}
\mathcal{L} = \frac{2C_{0}^3}{(2\pi\nu)^2}\frac{1}{\frac{\lambda_{v}+2\mu_{v}}{\rho}+\left(\frac{c_p}{C_v}-1\right)\frac{\kappa}{\rho c_p}},
\end{equation}
where $\lambda_v$ and $\mu_{v}$ are the first and second viscosity coefficients, $c_p$ and $c_v$ the specific heats at constant pressure and constant volume, and $C_0$=$\sqrt{(B/\rho)(c_p/C_v}$ with $B$ the water bulk modulus.
Indeed, intrinsic damping might hamper water oscillations frequencies in excess of the MWCNT fundamental mode and become severe when entering the THz acoustic range.

\section{Conclusions and perspectives}
The \textit{generation} mechanism of acoustic waves in water upon excitation of both a single-wall (SW) and a multi-wall (MW) CNT with laser pulses of temporal width ranging from 5 ns down to ps was theoretically investigated via a multi-scale theoretical approach encompassing analytical modelling, MD and FEM.
First we showed that, depending on the combination of CNT size and laser pulse duration, the CNT can act both as a thermophone and a mechanophone. 
The mechanophone is shown to play a significant role when the laser pulse duration is much shorter than the CNT characteristic cooling time towards the external water: $\tau\ll\tau_{th}$.
Second, we prove that activation of the mechanophone may enable the generation of few nanometers wavelength sound waves in water, matching the CNT acoustic frequencies. This is achieved for laser pulse duration shorter than the fundamental CNT period: $\tau<\mathit{\tau_{0}}$. A similar physics is expected to play a role in other nanosystems, as in the case of vertically grown gold capped nanowires on a solid substrate, where the role of the CNT is played by the substrate and that of water by the vertically grown nanowires \cite{gandolfi2022ultrafast} and in nanotube-water systems were the nanotube could be synthesized out of other 2D materials \cite{vialla2020time}.

The present work suggests an additional generation pathway on top of the one commonly addressed in the case of water-immersed single nano-objects excitated with ns laser pulses, where only the thermophone effect contributes. In addition to this, it points to perspective research to be undertaken to further progress in the field. Specifically, the effect of intrinsic damping, i.e. viscous and viscoelastic damping \cite{yu2021nanoparticle, yu2022energy}, a yet open issue in the present frequency range, will need to be addressed. Indeed, intrinsic damping might hamper water oscillations frequencies greater than that of the MWCNT fundamental mode. Upon further reducing the laser pulse duration, water might respond as a solid rather than a liquid, its dispersion becoming non linear and the sound velocity increasing to 3200 m/s \cite{ruocco2008history}. This would also enable to make a link with the elastic properties of ice in the hypersonic frequency range \cite{kuriakose2017longitudinal, sathyan2020three, sandeep20213d}. A systematic effect of the thermal boundary conductance in the generation of acoustic waves in liquid-immersed nanosystems, as the excitation laser pulse width is reduced, is still lacking. These aspects will bear great relevance in view of perspective applications.

\section*{Acknowledgements}
This work was supported by the LABEX iMUST (ANR-10-LABX-0064) of Université de Lyon, within the program "Investissements d'Avenir" (ANR-11-IDEX-0007) operated by the French National Research Agency (ANR), by ANR through project 2D-PRESTO under Reference ANR-19-CE09-0027, and through project ULTRASINGLE under Reference No. ANR-20-CE30-0016.\\
F.B. acknowledges financial support from CNRS through Délégation CNRS 2021-2022.
S.R. acknowledges financial support from the Italian Ministry of University and Research (PRIN Project QUANTUM2D).\\
M.F., F.M.B and A.C. thank the Politecnico di Torino’s High-Performance Computing Initiative (http://hpc.polito.it/) for the availability of computing resources.
\appendix
\section{Molecular dynamics}\label{MD}
Molecular dynamics simulations are adopted to: (i) calculate the TBC, $G$, at the SWCNT-water interface and (ii) show that water infiltrates the MWCNT in our configuration but not the SWCNT.\\

\noindent \textbf{TBC at the SWCNT-water interface}

The atomistic simulation protocol adopted to compute the TBC at the SWCNT-water interface is here described. We first obtain, from MD simulations, the temperature time dependence for the water-immersed SWCNT, $T_{cnt}^{md}(t)$, whose initial temperature has been increased with respect to that of the surrounding water. We then fit $T_{cnt}^{md}(t)$ with the numerical solution of the continuum model addressed in Section \ref{subsec:Thermal_dynamics}, $T(r,t; G)$, calculated for a coordinate $r$ inside the SWCNT wall and with $G$ as the fitting parameter.

Molecular dynamics simulations are implemented exploiting the LAMMPS package \cite{LAMMPS} while the initial set-up is created with the software Moltemplate \cite{Jewett2021} and VMD \cite{HUMP96}.
A Verlet integration scheme is used to integrate the equations of motion every fs for all simulations.

Long-range electrostatic interactions are simulated with a particle-particle particle-mesh solver (PPPM) \cite{Hockney1988}.
Carbon-carbon interactions are simulated by means of the reactive Tersoff force-field \cite{Tersoff1988, bellussi2021anisotropic}. 
Water molecules interactions are simulated with the extended version of the single point charge (SPC/E) force-field \cite{Berendsen1987}.
The rigid model for water bonds and angle is implemented in LAMMPS via the SHAKE algorithm \cite{RYCKAERT1977327}.
Carbon-oxygen interactions are simulated with a 12-6 Lennard-Jones potential with a cut-off radius of 1 nm.
Lennard-Jones parameters are taken with respect to the experimental contact angle of water and graphite, namely 86$^{\circ}$ \cite{Werder2003}.

A 10 nm long (5,5) SWCNT is placed in a 8.14 nm x 8.14 nm x 10.4 nm simulation box, with water molecules outside the cylindrical CNT cavity.
The system is first brought to equilibrium by means of an energy minimization with a conjugate gradient method; it is then simulated at 290 K and 1 bar in a NPT ensemble with a Nose-Hoover thermostat and barostat for 1 ns. 
Subsequently, another equilibration is performed in NVT with a Nose-Hoover thermostat at 290 K.
The temperature of the CNT only is then set at 360 K, while water is kept at 290 K for 200 ps with two different Nose-Hoover thermostats.
Finally, in a NVE ensemble, the temperature of the carbon nanotube (which is now let free to evolve), $T_{cnt}^{md}(t)$, and throughout water are monitored for 500 ps, which corresponds to the time needed to reach thermal equilibrium in the simulation domain. Fitting $T_{cnt}^{md}(t)$ to the numerical solution of the continuum thermal model yields $G$=2.86$\times$10$^{6}$ W/(m$^2$K). This value is also used in our modeling for the interface between MWCNT and water. A small difference in the $G$ value at the water interface could be expected between SWCNT and MWCNT, as shown in the case of graphene with single and multi-sheets \cite{alosious2020kapitza, alexeev2015kapitza}.\\

\noindent\textbf{Water infiltration in CNTs}

In the performed MD simulations water does not infiltrate the (5,5) SWCNT. However, repeating the simulations for a larger system, (10,10) SWCNT, spontaneous water infiltration is observed, the diameter of the tube being extremely important for the infiltration phenomena.
As we already said, direct MD simulations on the MWCNT were not performed. However, when it comes to the MWCNT investigated in this manuscript, its internal diameter exceeds that of a (10,10) SWCNT; thus, water infiltration is expected.

For these reasons, the (5,5) SWCNT simulated in the continuum model is considered hydrophobic, while the MWCNT is filled with water.
Further supporting this scenario is the fact that water entrance in MWCNT of similar dimensions has been observed by several techniques i.e. with neutron scattering \cite{Maniwa2005} and X-ray scattering \cite{Kolesnikov2004}.
\section{CNT Mechanical properties}\label{CNT Mechanics}
Devising a mechanical model for CNTs is a difficult task. As a matter of fact, CNTs are composed out of an  highly anisotropic material, graphite, which is rolled up and radially deformed to accommodate for the correct matching of the single atomic cylinders. 
In particular, the literature lacks of an ultimate and unanimous model for the  $c_{ij}$ coefficients of CNTs. Several approaches have been proposed \cite{lu1997elastic, velasco2009vibrations, zaeri2015elastic, palaci2005radial}, each one with its own merits and shortfalls.

For our purposes we proceed as follows. We model the CNT as a hollow cylinder made out of a homogeneous and isotropic material of Young modulus $E_Y$ and Poisson ratio $\nu$. The values of $E_Y$ and $\nu$ are adjusted to reproduce the correct frequencies of the breathing modes of the CNT. We note that this approach can lead to incorrect estimates for vibrational modes other than the breathing (for instance a lateral bending); nevertheless the model is simple and suits our purposes, while also retaining the symmetry of the problem.

In particular, for the SWCNT case, $E_Y$ and $\nu$ are calculated starting from the known 2D elastic properties of graphene, and correctly predict the experimentally observed frequency of the radial breathing mode reported in the literature. For the more complex MWCNT case, instead, we used experimental values found in the literature. Both cases are detailed in the next subsections.

\subsection*{SWCNT: elastic constants from an analytical model}

We consider a SWCNT as a rolled sheet of graphene, thus a possible approach consists in starting from its 2D Young modulus  $E^{2D}=340$ N/m ($E^{2D}= 1\,{\rm TPa} \times h$, where $1\,{\rm TPa}$ is the graphene planar Young modulus and $h=0.34\,{\rm nm}$ is the nominal thickness of the graphene plane) and Poisson ratio $\nu=0.15$. We note that graphene is also characterized by a (small) bending modulus, which {\em cannot} be calculated using the standard bulk elastic formulas for an isotropic material and rather depends on the elastic energy required to bend the atomic network of graphene and its in-plane bonds. Such a bending is clearly non-zero in a SWCNT but it does not change during the breathing oscillation, thus it will be neglected in the following.

When the radius of a SWCNT is expanded from its equilibrium value $r_0$ to $r=r_0+\Delta r$, the circumference $2\pi r$ of the rolled graphene layer is clearly stretched as well. Since we typically assume the SWCNT is long and unable to experience significant movements/contractions along its axis, we will consider that the SWCNT is fixed at infinity and that it {\em does not} expand {\em nor} contracts in the axial direction. Given this condition, graphene can only stretch along the tangential direction $\phi$ with a strain $\epsilon_{\phi \phi}=\Delta r / r_0$, while $\epsilon_{z z}=\epsilon_{z \phi}=0$. The stress tensor components of the cylindrical surface of the SWCNT can now be calculated using the constitutive strain-stress equations \cite{mase2009continuum}:
\begin{equation} \label{eq:strainstress}
\begin{split}
\epsilon_{\phi \phi}E^{2D}& =\sigma_{\phi \phi}- \nu \, \sigma_{zz} \\
\epsilon_{zz}E^{2D}& = \sigma_{zz} - \nu \, \sigma_{\phi \phi}=0\\
\epsilon_{z \phi}E^{2D}& =(1+\nu) \, \sigma_{z \phi}=0,\\
\end{split}
\end{equation}
where we neglected any strain or stress component in the direction perpendicular to the graphene plane, since they are not pertinent to the current 2D mechanic description; we note also that the unit of the planar tensor components $\sigma_{ij}$ is here ${\rm N/m}$. The second and third equations imply $\sigma_{zz}=\nu \sigma_{\phi \phi}$ and $\sigma_{z \phi}=0$. Thus, inserting these results in the first equation we obtain the stress and strain relations along the tangential directions
\begin{equation} \label{}
\begin{split}
\epsilon_{\phi \phi}= & \frac{1}{E^{2D}}\left(\sigma_{\phi \phi} - \nu^2  \sigma_{\phi \phi} \right)
=\frac{1-\nu^2}{E^{2D}} \sigma_{\phi \phi}=\frac{1}{\tilde{E}^{2D}} \sigma_{\phi \phi}
\end{split}
\end{equation}
where we defined an effective 2D Young modulus $\Tilde{E}^{2D}=E^{2D}/(1-\nu^2)$. This result has been derived just considering a model of the 2D elastic response of a graphene layer, which does not consider in any way the deformation of the graphene sheet in the radial direction: the atomic network is and remains one atom thick regardless the mechanical configuration of the SWCNT.

\subsubsection*{Calculation of the breathing frequency}

\begin{figure}[t!]
\centering
\includegraphics[scale=0.35]{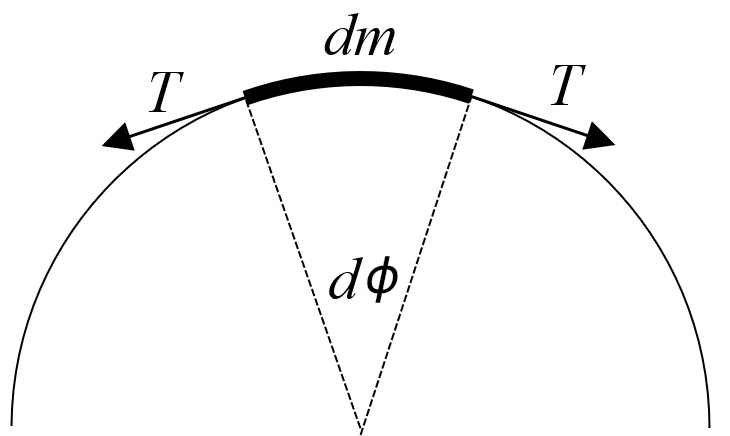}
\caption{Schematics of the tensions acting on an infinitesimal portion of the SWCNT.}
\label{SWCNT breathing mode}
\end{figure}

We now calculate the frequency of the breathing modes using the 2D elastic model sketched above. Let us assume we consider an infinitesimal portion of the SWCNT surface, of axial size $dz$ and azimuthal length $rd\phi$. The tangential tension acting on the edges of this cylindrical surface is $T = \sigma_{\phi\phi} dz$ and, given the curvature of the surface, it leads to a radial recovery force, see Fig. \ref{SWCNT breathing mode},
\begin{align}
    dF = T \, 2\sin (d\phi/2) \approx \sigma_{\phi\phi} d\phi dz. 
\end{align}

If we consider the mass of the cylindrical portion $dm=\rho  rd\phi dz$, where $\rho=0.76\,{\rm mg/m^2}$, the radial force will counteract the expansion of the SWCNT according to the equation
\begin{align}
    dm  \ddot{r} = (\rho rd\phi dz)\ddot{r}  = -dF = -\sigma_{\phi\phi} d\phi dz.
\end{align}

This yields the Ordinay Differential Equation (ODE) describing the radial breathing of the SWCNT, $\rho r \ddot{r} = - \sigma_{\phi\phi}$ or, considering that $r\approx r_0$,
\begin{align}
    \rho \frac{d^2}{dt^2}(r_0+\Delta r) =\rho\ddot{\Delta r} = -\frac{\sigma_{\phi\phi}}{r_0}=-\frac {\tilde{E}^{2D}}{r_0^2} \Delta r
\end{align}
and the expected radial oscillation frequency is
\begin{equation} \label{}
\omega = \sqrt{\Tilde{E}^{2D}/\rho r_0^2},
\end{equation}
which, in our (5,5) SWCNT case, gives $\omega=6.28\times10^{13}$ rad/s, in very good agreement with values found in literature \citep{velasco2009vibrations, lawler2005radial}. The same breathing frequencies can be reproduced by a bulk model of a $h$-thick cylinder shell made out of an isotropic material with a Young modulus $E_Y = \Tilde{E}^{2D} / h \approx 1$ TPa. $E_Y$ is precisely the Young modulus adopted in the main text to mimic the elastic constants of the SWCNT.

\subsection*{MWCNT: elastic constants from experimental values}
For the case of MWCNTs, we take the radial Young modulus $E_Y$ and Poisson ratio $\nu$, from published experimental data \citep{palaci2005radial}. In their work, Palaci et al. measured $E_Y$ for MWCNTs with a fixed ratio of their external and internal radii, $R_{ext}/R_{int}=2$. 
With reference to figure 3 of Palaci et al., $E_Y$ converges to 30 GPa for $R_{ext} >$ 4 nm. Our MWCNT follows the $R_{ext}/R_{int}=2$ prescription and has $R_{ext}=15.5$ nm. We then take $E_Y=30$ GPa and $\nu=0.28$ as prescribed in Palaci et al.

Remarkable for the sake of consistency, by extrapolating Palaci et al. Young modulus measurements for the number of walls decreasing to one, we fall on the Young modulus value obtained from our SWCNT model.

Inserting $E_Y$ and $\nu$ in equation \ref{elastic conversion} we then obtain the $c_{ij}$ coefficients for the CNT (both SW and MW). The result is a stiffness tensor $\textbf{C}$ of a homogeneous and isotropic material (automatically including cylindrical symmetry), which matches the real CNTs Young modulus and Poisson ratio in the radial direction. 

\bibliographystyle{model1-num-names}
\bibliography{cas-refs}

\end{document}